\documentstyle[prl,twocolumn,epsf,aps]{revtex}

\begin{document}
%
%
\newcommand{\msr}{$\mu$SR}


\twocolumn[\hsize\textwidth\columnwidth\hsize\csname@twocolumnfalse\endcsname

\title{ Electronic structure of the muonium center as a shallow donor in ZnO }

\author{Koichiro Shimomura, Kusuo Nishiyama, and Ryosuke Kadono\cite{soken}}
\address{Meson Science Laboratory,
Institute of Materials Structure Science,
High Energy Accelerator Research Organization (KEK),\\
1-1 Oho, Tsukuba, Ibaraki 305-0801, Japan}

\maketitle

\begin{abstract}
The electronic structure and the location of muonium centers (Mu) in
single-crystalline ZnO were determined for the first time. Two species of Mu
centers with extremely
small hyperfine parameters have been observed below 40 K. Both Mu centers
have an axial-symmetric hyperfine structure along with a $\langle$0001$\rangle$ axis,
indicating
that they are located at the ${\rm AB}_{O,\parallel}$ and ${\rm
BC}_{\parallel}$
sites.  It is inferred from their small ionization
energy ($\simeq 6$ meV and 50 meV) and hyperfine parameters ($\sim10^{-4}$
times the
vacuum value) that these centers behave as shallow donors, strongly
suggesting that hydrogen is one of the primary origins of
$n$ type conductivity in as-grown ZnO.
\end{abstract}

\pacs{61.72.Vv, 71.55.Gs, 76.75.+i}
]


 Zinc oxide (ZnO) is one of the most promising semiconductors for the next
generation of electronic and optelectronic devices. It has already been
applied to transducers, phosphors and varistors, due to its unique
piezoelectric, optical, and electrical properties. In these applications,
polycrystalline material has mainly been used. Moreover, recent progress in
single crystal growth \cite{dcl:98} has opened up new possibilities, like
bright blue and uv light emitters.  Optical uv lasing has already been
observed even at room temperature \cite{dmb:97}.

For applications to optoelectrical devices, it is crucial to control the
bulk electronic conductivity of crystalline ZnO.  However, it is notoriously
difficult to obtain intrinsic ZnO, ending up with materials showing strong
$n$ type conductivity.  In spite of more than 20 years of investigations,
the origin of this unintentional carrier doping is still controversial. It
has long been speculated that the dominant donor is a native defect, either
oxygen vacancy, or zinc interstitial \cite{ghe:90,dcl:99}.
Unfortunately, recent theoretical investigations have revealed that
none of those native defects behave as shallow donors \cite{afk:00}.

 Recently, it was theoretically pointed out that hydrogen (H), which is
quite difficult to remove from the crystal growth environment, is an excellent
candidate for such a shallow
 donor\cite{cgv:00}. As shown in Fig.~1, ZnO crystallizes in the wurtzite
structure corresponding to an elongated zinc blend structure with hexagonal
symmetry around the
 [0001] axis.  The lattice parameters are known by experiments as
$a=0.325$ nm, $c/a=1.602$ and $u=0.382$ in normalized coordinates. From a
first-principle calculation, the lowest energy configurations for hydrogen
are predicted to be at the BC$_{\perp}$ site, with a nearly equivalent
formation energy for the BC$_{\parallel}$, AB$_{O, \perp}$,
and
AB$_{O, \parallel}$ sites \cite{cgv:00}.  Experimental evidence
for this scenario has been claimed in several reports
\cite{emo:54,dgt:56,jil:57,sjb:97,kwk:94}, where an increase in the
conductivity was observed upon introducing H into ZnO.

   In this Letter we report on a determination of the electronic structure
and the location of a muonium (Mu, an analogue of isolated hydrogen whose
proton is substituted by a positive muon) as a shallow donor in ZnO. By
using single-crystalline ZnO, two species of muonium have been clearly
distinguished.  The
muonium center is readily observed in a wide variety of semiconductors
 after positive muon implantation, and has been serving as a unique
source of information
 on the electronic structure of {\sl isolated} hydrogen centers
\cite{bdp:88}.
 While the dynamical aspect (e.g., diffusion
 property) may be considerably different between Mu and H due to the light mass
of Mu
 ($\simeq\frac{1}{9} m_p$), the local electronic structure
 of Mu is virtually equivalent to that of H
 after a small correction due to the difference in the reduced mass
($\sim4$\%).
 It is now well established in elemental and III-V compound semiconductors
that there are two stable (and metastable) sites, one at the center of
the matrix bond (i.e., BC-site, Mu$_{BC}^{0}$) with a large outward
relaxation of the nearest-neighbor (nn) host atoms, and the other around the
center of a tetrahedron cage (i.e., $T_d$ site, Mu$_{T}^{0}$). While
Mu$_{T}^{0}$ has a large isotropic hyperfine parameter (almost the
same order of the vacuum value,
 $A_\mu=4463$ MHz), the hyperfine parameter of Mu$_{BC}^{0}$
has a value about one order of magnitude smaller with a large unpaired spin
density distributed
 on the nn host atoms.  Recently, a novel muonium state having an extremely
small hyperfine parameter ($10^{-4}\times A_\mu$) has been reported in a
II-VI compound semiconductor, CdS \cite{jmg:99}, suggesting that such a
shallow Mu center (and H center as well) might be present in ZnO to serve as
a donor.

  The experiment was performed at the Meson Science Laboratory (located in
KEK) which provides a pulsed (50 ns pulse width and 20 Hz repetition) beam
of 100\% spin-polarized muons with a beam energy of 4 MeV.  The muon beam
with longitudinal polarization was implanted into a single-crystalline 
wafer (40 mm diameter, 0.5mm
thickness, [0001] orientation) of ZnO
  obtained from Eagle-Picher Industries, Inc.
 The conventional time differential muon spin rotation (\msr)
measurements were performed under a magnetic field  applied in two different
orientations: one in the transverse direction (TF, $\vec{B}$ in Fig.~1) and the other
in a tilted direction ($\vec{B}'$) with respect to the initial muon spin polarization 
$\vec{P}_\mu$.
To obtain the tilted field, both transverse and longitudinal (LF) magnetic
fields were applied simultaneously.
  In the case of hyperfine parameter measurements, the specimen was placed
on a cold finger with the [0001] axis parallel
with $\vec{P}_\mu$ (O-face up). As shown in Fig.~1, the
[11$\bar{2}$0] axis was set either perpendicular to $\vec{B}$ (Fig.~1a) or parallel
with $\vec{B}$ (Fig.~1b)  to examine the angular dependence of the hyperfine constants.
For the temperature dependence measurements of the muonium
fraction, the [0001] axis was tilted by 45$^\circ$ to $\vec{P}_\mu$ 
(i.e.,$\vec{B}\angle[0001]=45^\circ$) while $\vec{B}\perp[11\bar{2}0]$. 

   It has been inferred from TF (=2.00 mT, 4.00 mT and 30.0 mT) measurements
that only a single diamagnetic muon state is present above 40 K.  The relaxation rate is almost
independent of temperature with a rate of $\simeq0.022(6)$ $\mu$s$^{-1}$ for
Gaussian damping, which is consistent with the dipole-dipole interaction of
muons with $^{67}$Zn nuclei (natural abundance 4.1\%).
On the other hand, the muon spin rotation signal changes drastically
below 40 K. A typical \msr\ spectrum 
is shown in Fig.~2 with fit errors, where the data were obtained under 30.0 mT
with $\vec{B}\angle[0001]=45^\circ$. 
Fig.~3 shows the temperature dependence of the FFT spectrum, in which two pairs of satellite lines
are seen with their position situated symmetrically around the central line corresponding to the
precession of diamagnetic muons (with the gyromagnetic ratio
$\gamma_\mu=2\pi\times135.53$ MHz/T) at 5 K.
The splitting of these satellites remained unchanged when the applied field
was changed to 2.00 mT or 4.00mT.
Moreover, a nearly equivalent frequency spectrum was observed when the specimen
was rotated by 90$^\circ$ around the [0001] axis (i.e., between (a) and (b) in Fig.~1).
These results strongly suggest that two muonium centers with
extremely small anisotropic hyperfine parameters exist in ZnO. The hyperfine
parameters are about $10^{-4}$  times smaller than the vacuum value and they
are symmetric to the [0001] axis.

Provided that the hyperfine interaction has an axial symmetry, we expect
two muonium precession signals for the high-field limits with frequencies
\begin{eqnarray}
\nu_-(\theta)&\simeq&\nu_0-\frac{1}{2}\Delta\nu(\theta),\\
\nu_+(\theta)&\simeq&\nu_0+\frac{1}{2}\Delta\nu(\theta),\\
\Delta\nu(\theta) &=& A(\theta) = \mid A_{\parallel}\cos^2(\theta) +
A_{\perp}\sin^2(\theta) \mid,
\end{eqnarray}
where $2\pi\nu_0=\gamma_\mu B$ with $B=|\vec{B}|$ being the applied field
($B\gg 2\pi A/\gamma_e$, where $\gamma_e=2\pi\times28.024$ GHz/T is the
gyromagnetic ratio of electron),  $\theta$ is the
angle between $\vec{B}$ and the symmetry axis [0001],
and $ A_{\parallel}$ and $ A_{\perp}$ are the hyperfine parameters parallel
and normal to [0001], respectively.
It was revealed upon preliminary analysis that the fitting of \msr\ time
spectra assuming the three frequency components  ($\nu_0$, $\nu_-$, and
$\nu_+$) did not reproduce the time spectrum, yielding a large fraction ($\sim20$ \%)
of fit errors and poor reduced $\chi^2$ ($\simeq2.40$). 
On the other hand, as suggested in Fig.~3a, fitting analysis with two sets of satellites
including five components ($\nu_0$, $\nu_{i-}$, and $\nu_{i+}$, $i=1,2$)
turned out to yield a satisfactory result with drastically improved $\chi^2$ ($\simeq1.52$). 
 This indicates that there are two species of Mu centers with respective fractional yields in this
compound.
 From the spectrum with $\vec{B}\perp[11\bar{2}0]$ (Fig.~1a), 
 the hyperfine parameters are deduced to be
\begin{eqnarray}
A _{1} (90^\circ) &=& \mid A_{1\perp}\mid = 358(4)\:{\rm kHz}, \\
A _{2} (90^\circ) &=& \mid A_{2\perp}\mid = 150(4)\:{\rm kHz}.
\end{eqnarray}
Combining this result with the data under a tilted field $\vec{B}'$ (where $\theta
=54.0^{\circ}, \Delta\nu_{1} = 495(2)$kHz, $\Delta\nu_{2} = 298(4)$kHz), the
rest of the
hyperfine parameters are deduced as
\begin{eqnarray}
\mid A_{1\:\parallel}\mid &=& 756(13)\:{\rm kHz}, \\
\mid A_{2\:\parallel}\mid &=& 579(19)\:{\rm kHz}.
\end{eqnarray}
As shown in TABLE I, the angular dependence of the frequencies
($\Delta\nu_{i}$ )
calculated by the above parameters is in excellent agreement with the
experimental observation.

   The possibility that these Mu centers have a hyperfine tensor with the
symmetry axis parallel to BC$_{\perp}$ is eliminated by the fact
that
the observed precession frequency is independent of the rotation of the crystal
 around the [0001]  axis by 90$^\circ$.
Another attempt to explain this by resorting to a sufficiently small
anisotropy with the symmetry axis parallel to BC$_{\perp}$
fails to account
for the difference between $A_{\parallel}$ and $A_{\perp}$  consistently
with the data.
Thus, we conclude that there are two species of Mu centers, both of which
have axially symmetric hyperfine structure along with the [0001] axis.
Hereafter, we denote these two centers as Mu$_{I}$ and
Mu$_{II}$
with the corresponding hyperfine parameters, $A_{1}(\theta)$ and
$A_{2}(\theta)$,
respectively.  The static dielectric constants in ZnO are reported to be
7.8(3) for perpendicular and 8.75(40) for parallel to the [0001] axis
\cite{dvd;01}.
The degree of obtained anisotropy for the muonium hyperfine tensor ($\sim$50\%)
is much larger than that of the dielectric constant ($\sim$10\%), indicating
that
the anisotropy is determined by the local electronic structure with
the BC$_{\parallel}$ and AB$_{O,\parallel}$ sites (see
Fig.~1) being the most probable candidates for the sites of those Mu centers.
Considering the magnitude of anisotropy in the hyperfine tensors,
it would be reasonable to presume that
Mu$_{I}$ is
  located at the AB$_{O, \parallel}$ site and Mu$_{II}$ at
 the BC$_{\parallel}$ site.
   Let us compare our results to a simple model of shallow
level centers in a dielectric medium. In this model,
the hyperfine parameter is inversely proportional to
the cube of the Bohr radius ($a_d$) of the bound electrons.
The isotropic part of the hyperfine parameter, $A_{\rm iso}$
($=\frac{1}{3}A_{\parallel}+\frac{2}{3}A_{\perp}$),
is 491 kHz for Mu$_{I}$ and 293 kHz for
Mu$_{II}$. Compared with $A_\mu=4463$ MHz,
one obtains $a_d=21a_{0}=1.1$ nm for Mu$_{I}$
and $a_d=25a_{0}=1.3$ nm for Mu$_{II}$ (where $a_0$
is the Bohr radius of the free Mu).  On the other hand, the Bohr
radius for a hydrogen-like defect is
calculated from the average dielectric constant, $\epsilon=8.12$,
and the electron effective mass, $m^{\star}=0.318m_{e}$, of ZnO \cite{dlr:75},
i.e. $a_d =(\epsilon/m_{e}/m^{\star})a_0=25.5a_0$.  This value is
qualitatively in
good accord with those of Mu$_{I}$ and Mu$_{II}$.

   The temperature dependence of the amplitudes of Mu$_{I}$,
Mu$_{II}$,  and diamagnetic muon are plotted in Fig.~4. The total
yield of all states are almost independent of temperature, suggesting that
Mu$_{I}$ and Mu$_{II}$ are ionized to a diamagnetic muon
above the transition temperature($\sim40$ K).
 It is unlikely that these Mu energy levels are just above the valence
band.  Otherwise, the temperature dependence of the muonium charge state
would not be expected
due to the $n$ type conductivity of the present specimen where the Fermi
level is much higher than the mid-gap level. These results indicate that the
Mu centers act  as shallow level donors. Thus, since Mu centers simulate the
electronic structure of H in ZnO,
our result provides convincing evidence that the hydrogen centers in ZnO are
shallow donors,
leading to $n$ type conductivity in ZnO.

The activation energies of Mu$_{I}$ and
Mu$_{II}$  were obtained to be 3 meV and 25 meV,
respectively from the data in Fig.~4.
According to the analysis in \onlinecite{jmg:99}, the relation $E_d=2E_a$ is
satisfied between the defect level energy ($E_d$) and the activation energy
($E_a$), which leads to the respective defect level energies of
Mu$_{I}$ and
Mu$_{II}$ to be 6 meV and 50 meV. The latter is
fairly consistent with the calculated value of the hydrogen-like impurity
model, $13.6(m^*/m_e/\epsilon^2)=66$ meV,
and the observed value of 61meV attributed to H in an earlier
report\cite{dcl:98}.
Considering the large ambiguity in determining the defect level energy for
another donor at 31 meV which has a much lower concentration  in
\onlinecite{jmg:99},
Mu$_{I}$ may correspond to this shallower donor.

   The reason for the absence of Mu centers at other interstitial sites is
yet to be understood. Another issue is that a large fraction of diamagnetic
muons (about 50\%) exists
even at the lowest temperature. One of the possibilities is that muon-oxygen 
bounding is formed, which has been commonly observed in various oxides.
The other is that the diamagnetic centers
may correspond to those at the BC$_\perp$ or AB$_\perp$ sites, where their
defect energy levels are in the conduction band and/or their hyperfine
parameters are too small to observe in our experiment. Further experiments,
including \msr\ measurements at different geometry, would be helpful to
address these issues.
Meanwhile, more accurate
theoretical investigations are strongly required to unambiguously identify
the observed Mu centers.

   In summary, we have demonstrated that two species of muonium centers are
formed in ZnO below 40 K with extremely small hyperfine parameters. These
centers have an axially symmetric hyperfine interaction around the [0001]
axis.
The temperature dependence of their fractional yields indicates that they
act as
shallow donors, strongly suggesting that hydrogen is the primary origin of
unintentional $n$ type conductivity in ZnO.

   We would like to thank the staff of KEK-MSL for their technical support.  In
particular, special thanks are due to K. Nagamine for his continuous
encouragement and support for this work.  We also appreciate helpful
discussions with S. Tsuneyuki on theoretical aspects of this work and
communications with G. Cantwell, D.C. Look and D. Eason on the bulk property
of the ZnO specimen.

Note added: After submission of this manuscript,
the presence of a muonium state in the powder sample of ZnO was reported
by a separate group\cite{cox:01}.

\newpage

\begin{figure}
\mbox{\epsfxsize=0.4\textwidth \epsfbox{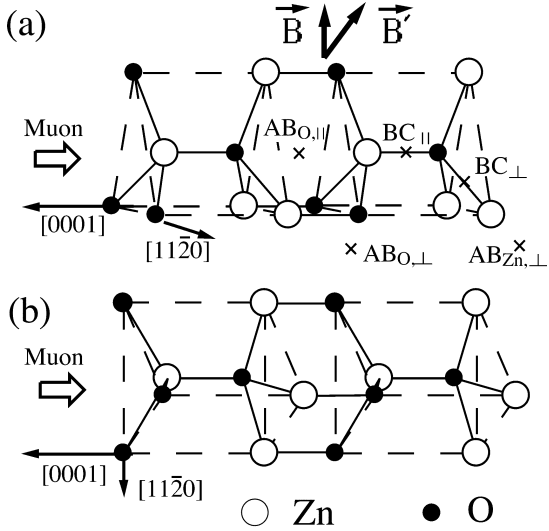}}      
\vspace{5mm}
\caption{The crystal structure of ZnO and the geometry of \msr\ measurements.
The [0001] axis is parallel with  the initial muon polarization $\vec{P}_\mu$, while
the transverse field $\vec{B}$ is either (a) perpendicular to or (b) parallel with
[11$\bar{2}$0] axis. Tilted field $\vec{B}'$ is in a plane defined by $\vec{B}$
and $\vec{P}_\mu$. ``BC" refers to
the bond center sites and ``AB" to the anti-bonding sites.}

\label{fig1}
\end{figure}

\begin{figure}
\mbox{\epsfxsize=0.4\textwidth \epsfbox{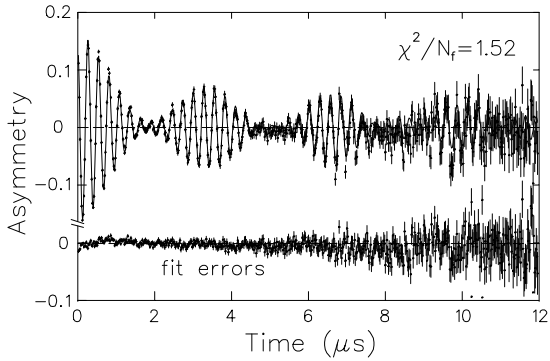}}      
\vspace{5mm}
\caption{\msr\ time spectrum in ZnO at 5.0 K, where the external field $\vec{B}$
($|\vec{B}|=30.0$ mT) was applied 45$^\circ$ to the [0001] axis.
A fitting result with two species of muonium centers are shown with
corresponding errors (difference between data and fitted curve).
}
\label{fig2}
\end{figure}

\begin{figure}
\mbox{\epsfxsize=0.4\textwidth \epsfbox{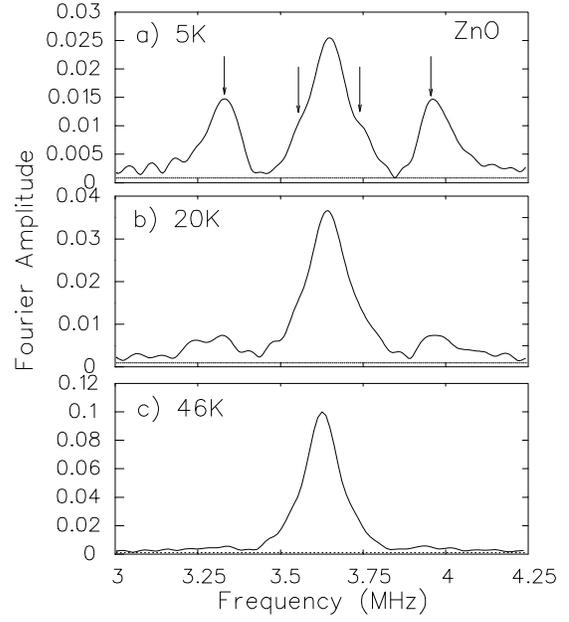}}      
\vspace{5mm}
\caption{FFT amplitude of the time spectrum at (a) 5.0 K, (b) 20 K, and
(c) 46 K, obtained with TF=30.0mT, where arrows indicate the satellite peaks.}
\label{fig3}
\end{figure}

\begin{figure}
\mbox{\epsfxsize=0.4\textwidth \epsfbox{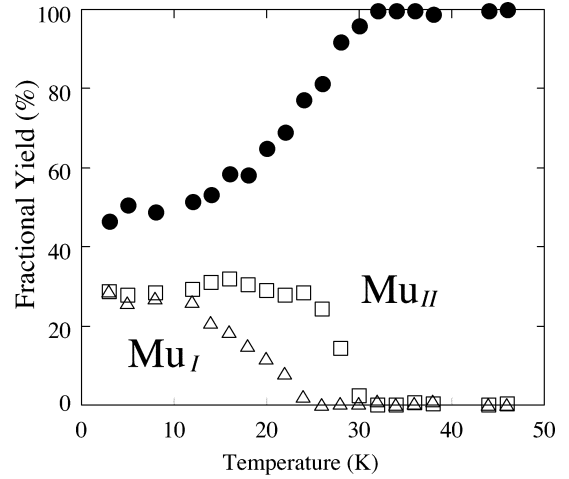}}      
\vspace{5mm}
\caption{The fractional yield of Mu$_{I}$ (open triangle),
Mu$_{II}$ (open square)  and a diamagnetic muon (closed circle)
versus temperature in ZnO. }
\label{fig4}
\end{figure}

\begin{table}
\caption{Hyperfine splitting of muonium centers in ZnO, where
the measured values are compared with those
calculated from a particular set of hyperfine parameters.}

\vspace{5mm}

\begin{tabular}{cccccc}
                   configuration            & $\theta$ & $\Delta\nu_{1cal}$  & $\Delta\nu_{1exp}$
 & $ \Delta\nu_{2cal}$  &$\Delta\nu_{2exp}$ \\
    \hline
$B\perp [11\bar{2}0]$ (Fig.1a) &  $90.0^{\circ}$  & 358(4) & 358(4) & 150(4) & 150(4) kHz  \\
$B\perp [11\bar{2}0]$ (Fig.1a) &  $54.0^{\circ}$  & 495(7) & 495(2) & 298(8) & 298(4) kHz  \\
$B\parallel [11\bar{2}0]$ (Fig.1b) &  $90.0^{\circ}$  & 358(4) & 356(2) & 150(4) & 153(4) kHz  \\
$B\angle [11\bar{2}0]=42.4^\circ$ (1b) &  $47.6^{\circ}$  & 539(8) & 541(4) & 345(9) & 350(4) kHz \\
\end{tabular}
\end{table}

\end{document}